\documentclass[11pt,twoside]{article}

\usepackage{asp2014}

\aspSuppressVolSlug
\resetcounters

\begin{document}

\title{FAIR high level data for Cherenkov astronomy }

\author{Mathieu~Servillat,$^1$ Catherine Boisson,$^1$ Matthias Fuessling,$^2$ and Bruno~Khelifi$^3$}
\affil{$^1$LUTH, Observatoire de Paris, Université PSL, Université de Paris, CNRS, F-92190 Meudon, France; \email{mathieu.servillat@obspm.fr}}
\affil{$^2$CTAO gGmbH, 69117, Heidelberg, Germany}
\affil{$^3$APC, IN2P3/CNRS - Université de Paris, 75013 Paris, France}

\paperauthor{Mathieu~Servillat}{mathieu.servillat@obspm.fr}{0000-0001-5443-4128}{LUTH - Observatoire de Paris}{}{Meudon}{}{92190}{France}
\paperauthor{Catherine Boisson}{catherine.boisson@obspm.fr}{0000-0001-5893-1797}{LUTH - Observatoire de Paris}{}{Meudon}{}{92190}{France}
\paperauthor{Matthias Fuessling}{matthias.fuessling@cta-observatory.org}{}{CTAO}{}{Heidelberg}{}{}{Germany}
\paperauthor{Bruno~Khelifi}{khelifi@apc.in2p3.fr}{0000-0001-6876-5577}{APC, IN2P3/CNRS - Université de Paris}{}{Paris}{}{75013}{France}



\begin{abstract}
We highlight here several solutions developed to make high-level Cheren-kov data FAIR: Findable, Accessible, Interoperable and Reusable. The first three FAIR principles may be ensured by properly indexing the data and using community standards, protocols and services, for example provided by the International Virtual Observatory Alliance (IVOA). However, the reusability principle is particularly subtle as the question of trust is raised. Provenance information, that describes the data origin and all transformations performed, is essential to ensure this trust, and it should come with the proper granularity and level of details. 

We developed a prototype platform to make the first H.E.S.S. public test data findable and accessible through the Virtual Observatory (VO). The exposed high-level data follows the gamma-ray astronomy data format (GADF) proposed as a community standard to ensure wider interoperability. We also designed a provenance management system in connection with the development of pipelines and analysis tools for CTA (ctapipe and gammapy), in order to collect rich and detailed provenance information, as recommended by the FAIR reusability principle. The prototype platform thus implements the main functionalities of a science gateway, including data search and access, online processing, and traceability of the various actions performed by a user.
\end{abstract}



\section{Introduction and context}

An effort is on-going to make science data FAIR: Findable, Accessible, Interoperable and Reusable \citep{Wilkinson2016}. In the astronomy domain, the International Virtual Observatory Alliance (IVOA, {\small\url{https://www.ivoa.net}}) provides standards since 2002 with similar objectives. We thus observe a connection between the FAIR principles and IVOA standards, demonstrating that, with these standards implemented, data management systems are close to compliance, with only a few extra components required to make astronomical data FAIR \citep{I4-001_adassxxxi}. 

Though the first three principles can be fulfilled with technical solutions, the reusability principle is more subtle because the question of trust is raised. It is stated that "data and metadata should be richly described with a plurality of accurate and relevant attributes". The wording is more qualitative than quantitative, indicating that the richness, the accuracy and relevance are to be adapted to each project. Detailed provenance of the provided data is indicated as essential to ensure this trust. An IVOA recommendation to model astronomy provenance metadata has been released in 2020 as to set up the basics to structure provenance information \citep{2020ivoa.spec.0411S}. The provenance should then be provided at the appropriate granularity and level of detail for each project. In order to make provenance of astronomy data more practical, we organized discussions on common approaches across all wavebands with current and future large observatories within the ESCAPE European project ({\small\url{https://www.projectescape.eu}}), the IVOA and at ADASS \citep{B9-56_adassxxx}.

Cherenkov astronomy provides an interesting use case, as complex data is acquired and processed based on the detection of Cherenkov light generated in the atmosphere by air showers induced by energetic cosmic particles. Unlike current experiments such as H.E.S.S., MAGIC or VERITAS, the next generation Cherenkov Telescope Array (CTA) will be an open observatory providing a public access to its high level science data products, with the requirement to provide FAIR data. The complex treatment of raw Cherenkov data thus implies to attach a detailed description of the high level data and their provenance, so as to explain the different assumptions made during the data processing, and building users' trust.

We developed a prototype platform to expose the first H.E.S.S. public test data release through the Virtual Observatory, making it findable and accessible (see \S\ref{F-A}). This high-level data follows a proposed standard to ensure wider interoperability (see \S\ref{interoperability}). Along with this standardization, the development of dedicated Science Tools makes the data even more accessible and interoperable to end users (see \S\ref{gammapy}). We finally designed a provenance management system in connection with the development of pipelines and analysis tools for CTA (ctapipe and gammapy), in order to collect rich and detailed provenance information (see \S\ref{provenance}). The prototype platform thus implements the main functionalities of a science gateway, including data search and access, online processing, and traceability of the various actions performed by a user.


\section{VO access to H.E.S.S. public test data}
\label{F-A}

A collection of data has been publicly released\footnote{\url{ https://www.mpi-hd.mpg.de/hfm/HESS/pages/dl3-dr1}} by the H.E.S.S. collaboration in 2018. The released data contains lists of candidate gamma-ray photons detected by H.E.S.S. during runs of observations on specific targets, along with instrumental response functions. This high-level data (defined as data level 3, DL3), has already been specifically processed from raw data and serves as a base for further analysis by the end users.

We created a VO access to this H.E.S.S. DL3 public data\footnote{\url{ https://hess-dr.obspm.fr}}, for which we adapted the IVOA observation core metadata \citep[ObsCore,][]{2017ivoa.spec.0509L} in order to make Cherenkov data findable and accessible. The service can be queried using the IVOA Table Access Protocole (TAP) to further explore the available data. The resulting IVOA ObsTAP service is declared to the VO registry, so that it can be found by VO tools (such as Aladin or Topcat).

\section{A common format for interoperability}
\label{interoperability}

In gamma-ray astronomy, a variety of data formats and proprietary software have been traditionally used, often developed for one specific mission or experiment. For ground-based imaging atmospheric Cherenkov telescopes (IACTs), data and software are mostly private to the collaborations operating the telescopes. 
To foster interoperability, the community is developing high-level data format specifications through the project Gam-ma-ray Astronomy Data Format\footnote{\url{ https://gamma-astro-data-formats.readthedocs.io}} (GADF, \citealt{Deil2017}). This effort is necessary to ensure a wider use of the data, but it also helps to build confidence in the data.

\section{The open-source science tool gammapy}
\label{gammapy}

Gammapy\footnote{\url{ https://gammapy.org}} is an open-source Python package for gamma-ray astronomy built on Num-py, Scipy and Astropy \citep{GammapyI2017ICRC...35..766D,GammapyII2019A&A...625A..10N}. It is used as core library for the Science Analysis tools of the Cherenkov Telescope Array, is the recommended tool by the H.E.S.S. collaboration to be used for Science publications, and is already widely used in the analysis of existing gamma-ray instruments, such as MAGIC, VERITAS and HAWC.
The development of gammapy is carried out taking into account most of the FAIR4RS principles \citep{FAIR4RS}, in order to improve the software productivity, quality, reproducibility, and sustainability. This effort again promotes interoperability and builds confidence in data analysis.


\section{Providing provenance to foster reusability}
\label{provenance}

Provenance information has been discussed early within the IVOA\footnote{\url{https://wiki.ivoa.net/twiki/bin/view/IVOA/ProvenanceDataModelLegacy}}. Because of the complexity of Cherenkov data, CTA served as a major use case for the development of the IVOA Provenance data model \citep{2019ASPC..521..469S}. This served to further develop a provenance management system adapted to astronomical projects needs \citep{Servillat_ProvWeek2021}.

Additional use cases have been discussed within the IVOA and the European ESCAPE project \citep{B9-56_adassxxx}. We observed that some projects already have data collections generated and archived from which the provenance has to be extracted (provenance "on top"). Other projects, such as the CTA project, are building complex pipelines that can be used to test the automatic capture of the provenance information (capture "inside"). Different tools and prototypes have been developed and tested to capture, store, access and visualize the provenance information, which participate to the shaping of a full provenance management system able to handle detailed provenance information.

Provenance graphs can become extremely complex. For simpler use, the concept of a last-step flat provenance has recently been proposed: it is information on the last activity (execution and software description) and the context (workflow, instrument), that can be embedded into an entity as a list of keywords. It is then important to include the list of identifiers of generated and used entities, so that a full provenance may be reconstructed from this last-step provenance.

\acknowledgements We acknowledge support from the ESCAPE project funded by the EU Horizon 2020 research and innovation program (Grant Agreement n.824064). Additional funding was provided by the INSU (Action Sp\'ecifique Observatoire Virtuel, ASOV), the Action F\'ed\'eratrice CTA at the Observatoire de Paris and the Paris Astronomical Data Centre (PADC).

\bibliography{O4-002}  


\end{document}